\def\dir{./}
\documentclass[useAMS,usenatbib]{\dir mn2e}
\usepackage[fleqn]{amsmath}
\usepackage{amssymb}
\usepackage[super]{nth}
\usepackage{natbib}
\usepackage{graphicx}
\usepackage{float}
\usepackage{mathrsfs}
\usepackage{mathtools}
\usepackage{multirow}
\usepackage[usenames,dvipsnames]{color}
\usepackage{\dir aas_macros}
\usepackage{comment}
\usepackage{subfigure}
\usepackage{hyperref}
\usepackage{mathrsfs}
\usepackage{epstopdf, dcolumn}
\DeclareGraphicsExtensions{.pdf}
\usepackage{wasysym}
\usepackage{dashrule}
\usepackage{xcolor}
\usepackage{threeparttable}

\newlength\replength
\newcommand\repfrac{.33}

\newcommand\rulewidth{.6pt}
\newcommand\tdashfill[1][\repfrac]{\cleaders\hbox to \replength{%
  \smash{\rule[\arraystretch\ht\strutbox]{\repfrac\replength}{\rulewidth}}}\hfill}

\newcommand\tdotfill[1][\repfrac]{\cleaders\hbox to \replength{%
  \smash{\raisebox{\arraystretch\dimexpr\ht\strutbox-.1ex\relax}{.}}}\hfill}

\newcommand{\Rom}[1]{\uppercase\expandafter{\romannumeral #1}}
\newcommand{\rom}[1]{\lowercase\expandafter{\romannumeral #1}}
\newcommand{\zf}{\textit{ZF-COSMOS-20115}}
\newcommand{\Msol}{\textrm{M}_\odot}
\newcommand{\meraxes}{\textsc{Meraxes}}
\newcommand{\tiamat}{\textit{Tiamat-125-HR}}
\newcommand{\tocf}{21\textsc{cmFAST}}

\definecolor{colorBHoff}{RGB}{228,26,28}
\definecolor{colorBHon}{RGB}{55,126,184}
\definecolor{colorBHonstrong}{RGB}{77,175,74}

\title[DRAGONS \Rom{13}: ZF-COSMOS-20115 analogues in DRAGONS]{\LARGE{Dark-ages~Reionization~and~Galaxy~Formation~Simulation~-~\Rom{13}.}\\ AGN quenching of high-redshift star formation in {\zf}}

\author[Qin et al.]{Yuxiang Qin$^{1}$\thanks{E-mail: Yuxiang.L.Qin@gmail.com},
	Simon J. Mutch$^1$, Alan R. Duffy$^2$, Paul M. Geil$^1$, 
	\newauthor Gregory B. Poole$^2$, Andrei Mesinger$^3$ and J. Stuart B. Wyithe$^1$\thanks{E-mail: swyithe@unimelb.edu.au}\\
	$^{1}$School of Physics, University of Melbourne, Parkville, VIC 3010, Australia\\
	$^{2}$Centre for Astrophysics and Supercomputing, Swinburne University of Technology, PO Box 218, Hawthorn VIC 3122, Australia\\
	$^{3}$Scuola Normale Superiore, Piazza dei Cavalieri 7, I-56126 Pisa, Italy}

\begin{document}

\date{Accepted 2017 July 19. Received 2017 July 19; in original form 2017 April 11}
\pagerange{\pageref{firstpage}--\pageref{lastpage}} \pubyear{2017}
\maketitle
\label{firstpage}

\begin{abstract}
Massive quiescent galaxies (MQGs) are thought to have formed stars rapidly at early times followed by a long period of quiescence. The recent discovery of a MQG, {\zf} at $z\sim4$, only 1.5 Gyr after the big bang, places new constraints on galaxy growth and the role of feedback in early star formation. Spectroscopic follow-up confirmed {\zf} as a MQG at $z=3.717$ with an estimated stellar mass of ${\sim}10^{11}\mathrm{M}_\odot$, showing no evidence of recent star formation. We use the {\meraxes} semi-analytic model to investigate how {\zf} analogues build stellar mass, and why they become quiescent. We identify three analogue galaxies with similar properties to {\zf}. We find that {\zf} is likely hosted by a massive halo with virial mass of ${\sim}10^{13}\Msol$, having been through significant mergers at early times. These merger events drove intense growth of the nucleus, which later prevented cooling and quenched star formation. {\color{black}Therefore, {\zf} is unlikely to have experienced strong or extended star formation events} at $z<3.7$. We find that the analogues host the most massive black holes in our simulation and were luminous quasars at $z\sim5$, indicating that {\zf} and other MQGs may be the descendants of high-redshift quasars. In addition, the model suggests that {\zf} formed in a region of intergalactic medium that was reionized early.
\end{abstract}


\section{Introduction}
Massive quiescent galaxies (MQGs) are galaxies with stellar masses of the order of $10^{11}\Msol$, and {\color{black}low or null star formation rate (SFR).} It is thought that these objects formed rapidly at early times, followed by a long state of quiescence \citep{Ilbert2013}. Due to the absence of recent star-forming events in MQGs, they are observationally challenging to study, especially at high redshift. Using the ultra deep imaging from the FourStar Galaxy Evolution Survey\footnote{\url{http://zfourge.tamu.edu/}} (ZFOURGE), \citet{Straatman2014} identified 15 MQGs at $z=3.4-4.2$ with an estimated number density of MQGs (hereafter the S14 sample) of $(1.8\pm0.7)\times10^{-5} \mathrm{Mpc}^{-3}$. Despite the large uncertainties, this number density is significantly higher than the value expected from extrapolations using observations at lower redshifts \citep{Bell2003,Muzzin2013}, leading to an unexpectedly high MQG fraction of $34\pm13$ per cent at $z\sim4$.

Recently, spectroscopic follow-up \citep{Glazebrook2017} was performed on the brightest MQG in the S14 sample\footnote{The ID of {\zf} is 13172 in the S14 sample.}, {\zf}. Their analysis revealed that {\zf} has a stellar mass of ${M_*=1.7^{+0.12}_{-0.24}\times10^{11}\Msol}$ {\color{black}(see a more recent study of \citealt{Simpson2017})}, and redshift of $3.717\pm0.001$, but no detectable ongoing star formation based on Balmer absorption lines (i.e. $\textrm{H}\beta$, $\textrm{H}\gamma$ and $\textrm{H}\delta$), implying a current $\mathrm{SFR}<0.2\Msol\mathrm{yr}^{-1}$. Furthermore, modelling the spectral evolution of the stellar population \citep{1999astro.ph.12179F} suggests rapid growth of stellar mass with a $\mathrm{SFR}>990\Msol\mathrm{yr}^{-1}$ at the peak of activity and a formation time-scale less than 250 Myr.

In this paper we use the {\meraxes} semi-analytic model \citep{Mutch2016a} within the Dark-ages Reionization And Galaxy formation Observables from Numerical Simulations (DRAGONS\footnote{\url{http://dragons.ph.unimelb.edu.au}}) programme to investigate the galaxy formation history of MQGs at high redshift. We note that cosmological simulations have had difficulty producing MQGs at high redshift \citep{Lee2013,Wellons2015,Dave2016,Behroozi2016}. However, we identify three galaxies with properties similar to {\zf}.

This paper is organized as follows. We begin with a brief overview of the DRAGONS framework in Section~\ref{sec:dragons} and present the modelled galaxy property in Section \ref{sec:GMF}. We show the {\zf} analogues in our model in Section~\ref{sec:analogues} and discuss the history and future of {\zf} in Sections~\ref{sec:history} and \ref{sec:future}. Conclusions are given in Section \ref{sec:conclusions}. In this work, we adopt cosmological parameters from the Planck 2015 results ($\Omega_{\mathrm{m}}, \Omega_{\mathrm{b}}, \Omega_{\mathrm{\Lambda}}, h, \sigma_8, n_\mathrm{s} $ = 0.308, 0.0484, 0.692, 0.678, 0.815, 0.968; \citealt{PlanckCollaboration2015}). 

\section{DRAGONS}\label{sec:dragons}
The {\meraxes} semi-analytic model \citep{Mutch2016a} was specifically designed to study galaxy formation at high redshift and the epoch of reionization (EoR). In this work, we use the updated version of {\meraxes} \citep{Qin2017}, which includes a detailed prescription of black hole growth and AGN feedback. Our fiducial model was run on the {\tiamat} dark matter halo merger trees. We briefly review the model here and refer the interested reader to the aforementioned references for details.

The underlying {\tiamat} dark matter halo merger trees provide 100 snapshots between $z=35$ and 5 with a time interval of ${\sim}11.1$ Myr and 114 additional snapshots between $z=5$ and {\color{black}0.56} separated equally in units of Hubble time (Poole et al. in preparation). The particle mass resolution of {\tiamat} is ${\sim} 1.2\times10^8h^{-1}\Msol$ and the box size is $125h^{-1}$Mpc. In order to show convergence, {\meraxes} was also run using the \textit{Tiamat} halo merger trees \citep{Poole2015}, which were constructed from an N-body simulation sharing identical cosmology with {\tiamat} but at a higher mass resolution ($2.6\times10^{6}h^{-1}\Msol$) and smaller volume ($67.8h^{-1}$Mpc). These results are presented in Section \ref{sec:history} for comparison.

\begin{figure*}
	\begin{minipage}{\textwidth}
		\centering
		\vspace*{-3mm}
		\includegraphics[width=0.33\textwidth]{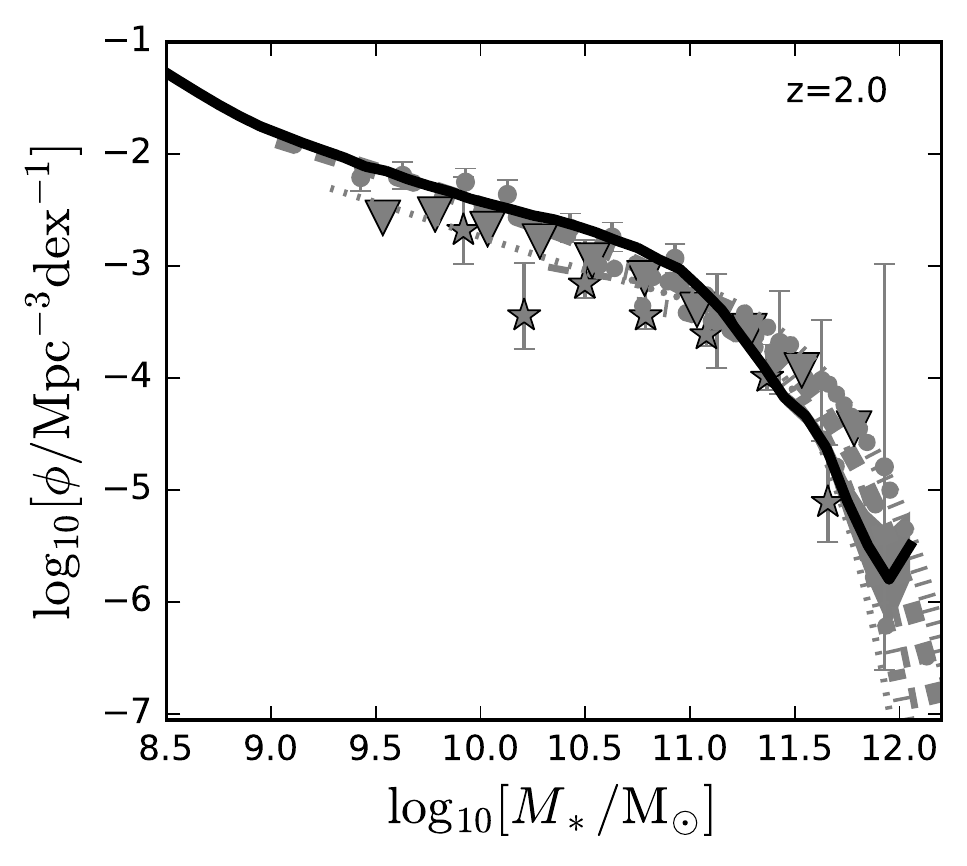}
		\includegraphics[width=0.33\textwidth]{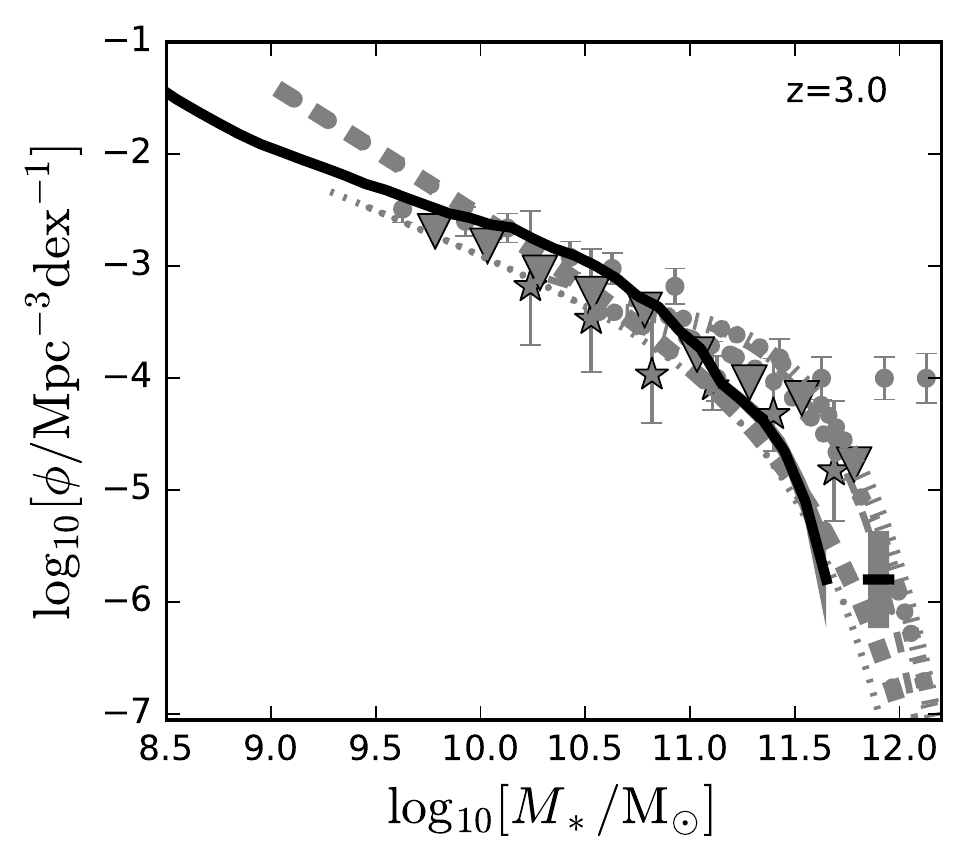}
		\includegraphics[width=0.33\textwidth]{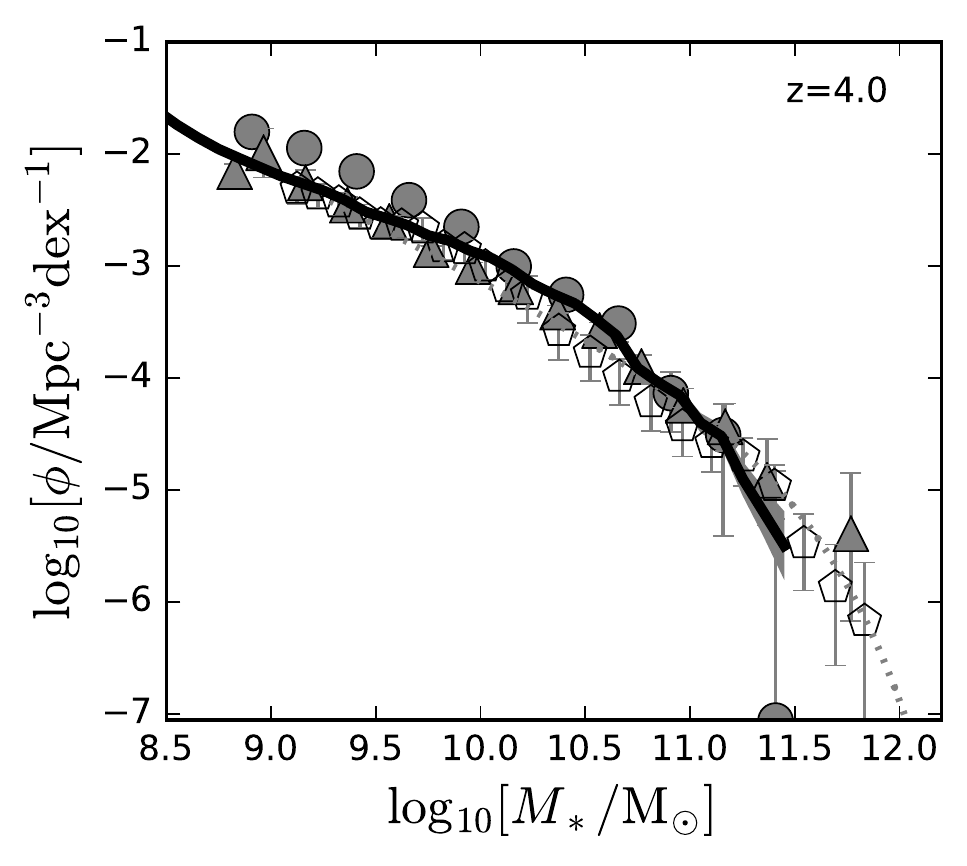}\\		\vspace*{-2mm}
		\includegraphics[width=0.33\textwidth]{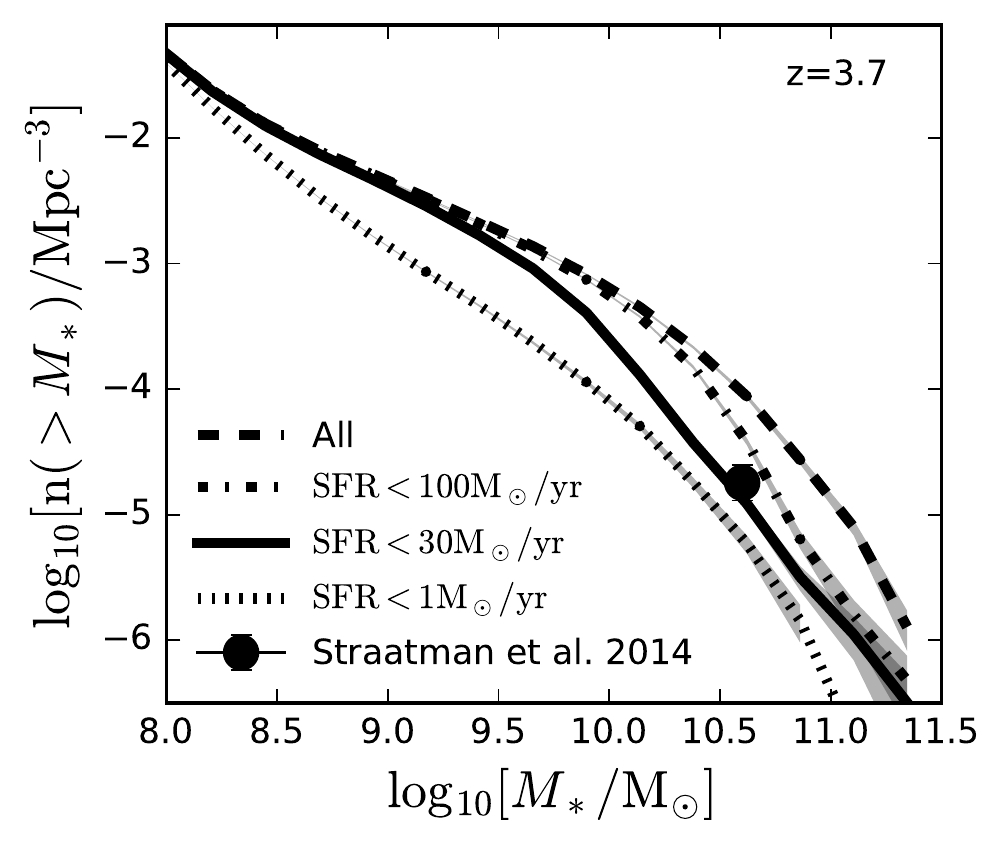}
		\includegraphics[width=0.33\textwidth]{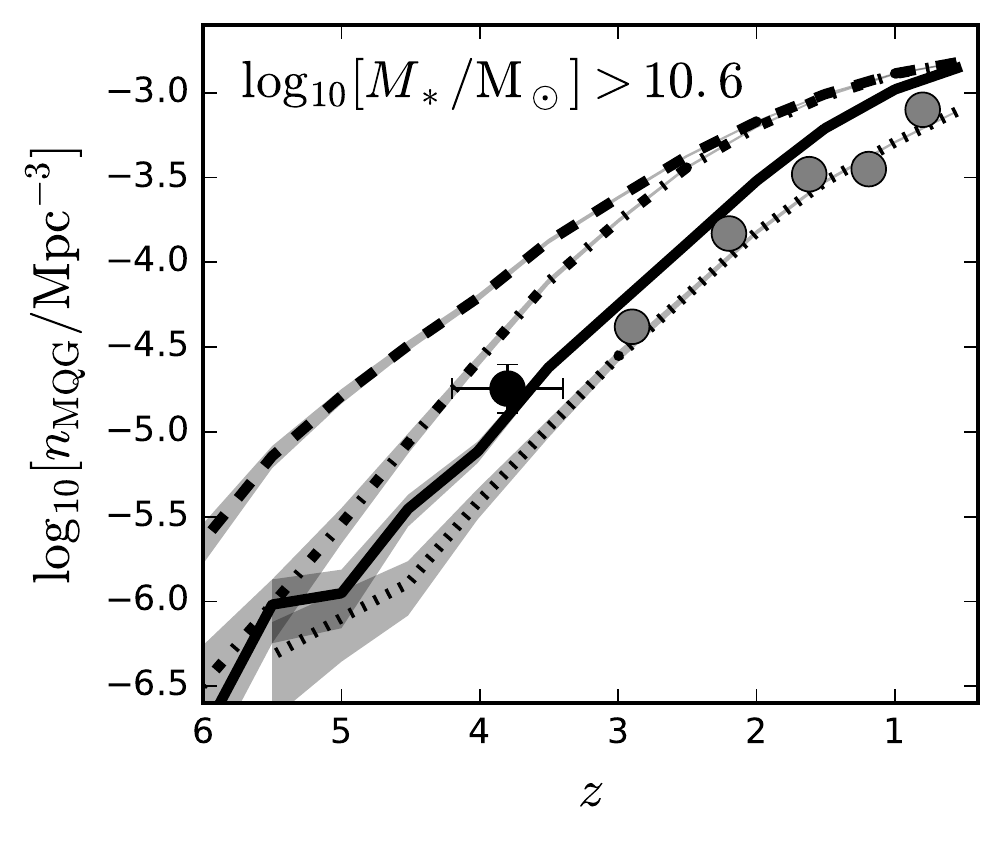}
		\includegraphics[width=0.33\textwidth]{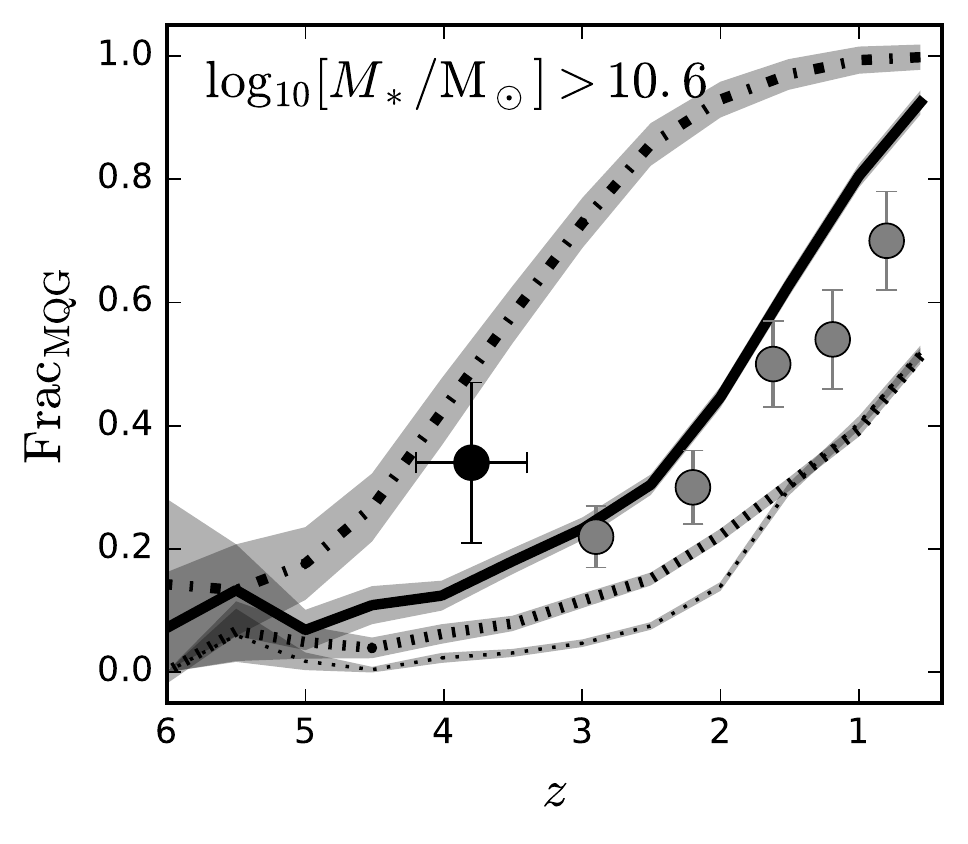}
	\end{minipage}
	\caption{\label{fig:gsmf} \textit{Top panels:} galaxy stellar mass function at $z=2-4$ compared to the observational data sets ({\color{gray}{$\bigstar$}}\citealt{Marchesini2009ApJ...701.1765M}, {\color{gray}{$\bullet$}}\citealt{Mortlock2011MNRAS.413.2845M}, {\color{gray}{{\hdashrule[0.6mm]{6mm}{3pt}{1.5mm 0.5mm}}}}\citealt{Santini2012}, {\color{gray}{\hdashrule[0.6mm]{6mm}{1.5pt}{1.5mm 0.5mm}}}\citealt{Ilbert2013}, {\color{gray}{\hdashrule[0.6mm]{6mm}{3pt}{1.5mm 0.4mm 0.4mm 0.4mm}}}\citealt{Muzzin2013}, {\color{gray}{$\blacktriangledown$}}\citealt{Tomczak2014}, {\color{gray}{\CIRCLE}}\citealt{Duncan2014}, {\color{gray}{$\blacktriangle$}}\citealt{Grazian2015}, {\color{gray}{\hdashrule[0.6mm]{6mm}{3pt}{0.4mm 0.4mm 0.4mm 0.4mm}}}\citealt{Huertas-Company2016}, {\color{gray}{$\pentagon$}}\citealt{Stefanon2016arXiv161109354S}, {\color{gray}{\hdashrule[0.6mm]{6mm}{1.5pt}{0.4mm 0.4mm 0.4mm 0.4mm}}}\citealt{Davidzon2017}). The shaded regions represent the 1$\sigma$ Poisson uncertainties. \textit{Bottom left panel:} number density of galaxies with masses above $M_*$ at $z=3.7$. Dashed, dash-dotted, solid and dotted thick lines represent samples of i) all galaxies; ii) galaxies with $\mathrm{SFR}<100\Msol\mathrm{yr}^{-1}$; iii) $\mathrm{SFR}<30\Msol\mathrm{yr}^{-1}$ and iv) $\mathrm{SFR}<1\Msol\mathrm{yr}^{-1}$. The black circle indicates the observationally estimated value and uncertainties from the S14 sample \citep{Straatman2014}. \textit{Bottom middle panel:} evolution of the number density of massive galaxies with $\log_{10}[M_*/{\rm M}_\odot]>10.6$. The observational data at lower redshifts ($z<3$; \citealt{Straatman2015}) is indicated with grey circles. \textit{Bottom right panel:} evolution of the MQG fraction. The MQG fraction with $\mathrm{SFR}<1\Msol\mathrm{yr}^{-1}$ of the model running with Eddington ratio, $\epsilon=0.1$ is shown with the dotted thin line for comparison.}
\end{figure*}

The {\meraxes} semi-analytic model consists of a number of important astrophysical processes, including gas infall, cooling, star formation, supernova feedback, AGN feedback, metal enrichment, mergers and reionization \citep{Mutch2016a,Qin2017}. The model was calibrated against the observed stellar mass function at $z\sim7-{\color{black}0.56}$ (see the galaxy stellar mass function at $z=2-4$ in the top panels of Fig. \ref{fig:gsmf}), black hole mass function at $z\lesssim0.5$, ionizing emissivity at $z\sim5-2$ and the Thomson scattering optical depth\footnote{The model running with the {\tiamat} merger trees is limited by the resolution and underestimates the ionizing emissivity at higher redshifts (see more in \citealt{Qin2017}).}. The model is in agreement with the observed Magorrian relation at $z<{\color{black}0.56}$ as well as the observed quasar luminosity function across a large redshift range ($z\sim6-{\color{black}0.56}$) when an opening angle of 80 deg is chosen to account for obscuration by dust. 

{\color{black} Compared to other simulations (e.g. the Millennium simulation; \citealt{Rong2017}),} several features of this model are well suited to study of high-redshift MQGs. We note that the cadence of our model is about 11 Myr at $z\ge5$ and reaches 30 Myr at $z=3.7$. This high temporal resolution enables us to not only resolve the dynamical time of the disc and simulate the bursty nature of star formation events, but also investigate the evolution of galaxies in more detail. In addition, the agreement between our model and observations, including the stellar mass function at $z\sim7-{\color{black}0.56}$ and quasar luminosity function at $z\sim6-{\color{black}0.56}$, suggests that the model galaxy and AGN catalogues we have constructed are able to represent the relevant observables across cosmic time. This is crucial for investigation of individual galaxy analogues within a cosmological context \citep[e.g.][]{Mutch2016b,Waters2016}.

\begin{table*}
	\caption{A list of the properties of the {\zf} analogues at $z\sim3.7$ and the time of peak star formation. The estimated properties of {\zf} by \citet{Glazebrook2017} are shown for comparison.}
	\begin{threeparttable}
		\label{tab:properties}
		\begin{tabular}{l|c|c|c|c|c}
			\hline \hline
			Name & $M_*(z=3.7)\mathrm{M}_\odot$ & $\mathrm{SFR}(z=3.7)/\mathrm{M}_\odot\mathrm{yr}^{-1}$ & $M_\mathrm{vir}(z=3.7)/\mathrm{M}_\odot$ & $z_\mathrm{peak}$ & $\mathrm{SFR}(z=z_\mathrm{peak})/\mathrm{M}_\odot\mathrm{yr}^{-1}$\\
			\hline
			ZF-1&{\color{black}$8.92\times10^{10}$}&0.0&$6.43\times10^{12}$&5.1&{\color{black}574.2} \\
			ZF-2&{\color{black}$1.02\times10^{11}$}&0.0&$5.46\times10^{12}$&4.9&{\color{black}827.2} \\
			ZF-3&{\color{black}$9.38\times10^{10}$}&0.0&$1.12\times10^{13}$&5.7&{\color{black}1000.1} \\
			{\zf}\tnote{a}&$(1.46-1.82)\times10^{11}$&$<0.2\ (<4\ \mathrm{from\ H\beta})$&$3.0\times10^{12}$&$5-8$&$>990$\\
			\hline
		\end{tabular}
		\begin{tablenotes}
			\item[a] A brief overview of how the {\zf} properties were estimated:
			\begin{enumerate}
				\item $M_*(z=3.7)$ from SED fitting;
				\item $\mathrm{SFR}(z=3.7)$ from Balmer lines;
				\item $M_\mathrm{vir}(z=3.7)$ from galaxy number density and halo mass function;
				\item $z_\mathrm{peak}$, $\mathrm{SFR}(z=z_\mathrm{peak})$ from spectral evolution of stellar population modelling. 
			\end{enumerate}
		\end{tablenotes}
	\end{threeparttable}
\end{table*}

\section{Galaxy mass function}\label{sec:GMF}

In the top panels of Fig. \ref{fig:gsmf} we plot the evolution of the model galaxy stellar mass function between $z=2$ and 4. The agreement with observed mass functions demonstrates the ability of the model to reproduce the observed growth of stellar mass across the broad range of cosmic time relevant to this work \citep{Qin2017}. The bottom left panel of Fig. \ref{fig:gsmf} presents the predicted cumulative galaxy stellar mass function at $z=3.7$ for four different samples in our model: i) all galaxies; ii) $\mathrm{SFR}<100\Msol\mathrm{yr}^{-1}$; iii) $\mathrm{SFR}<30\Msol\mathrm{yr}^{-1}$ and iv) $\mathrm{SFR}<1\Msol\mathrm{yr}^{-1}$. We note that there are 15 MQGs detected in the S14 sample, corresponding to a number density of $(1.8\pm0.7)\times10^{-5} \mathrm{Mpc}^{-3}$. This is shown in Fig. \ref{fig:gsmf} for comparison. In addition, SFRs derived from stellar population modelling vary from 0 to ${\sim}32\Msol\mathrm{yr}^{-1}$ in the S14 sample. We see that the number density of MQGs predicted by our model with SFRs in this range is in good agreement with the S14 sample. We also show the evolution of the number density of massive galaxies and the MQG fraction in Fig. \ref{fig:gsmf}. The model is consistent with observations at lower redshifts ($z<3$; \citealt{Straatman2015}). 

We note that, despite the large uncertainty, observations may suggest that the MQG fraction stops decreasing at $z\sim4$ while a continuously decreasing trend towards higher redshifts is usually presented in theoretical models (\citealt{Glazebrook2017}; see the bottom right panel of Fig. \ref{fig:gsmf}). {\color{black}The MQG fraction in the S14 sample is $34\pm13$ per cent at $z\sim4$.} This was compared to Illustris \citep{Vogelsberger2014}, a suite of hydrodynamical simulations of galaxy formation. Only ${\sim}20$ per cent of massive galaxies in Illustris are MQGs (in agreement with our model, see the $\mathrm{SFR}<30\Msol\mathrm{yr}^{-1}$ line in the bottom right panel of Fig. \ref{fig:gsmf}). Based on this, \citet{Straatman2015} suggested that quenching is likely to occur at early times in massive galaxies. However, we note that only one object in the S14 sample, {\zf}, has been spectroscopically confirmed to be a MQG and, due to the limited volume of the ZFOURGE survey and potential contaminations from dust obscured star-forming galaxies \citep{Straatman2015}, the MQG population at high redshift is not well constrained. Therefore, instead of making attempts to interpret the high MQG fraction at high redshift, in this work, we focus on finding analogues of the spectroscopically confirmed {\zf} and discuss the possible evolutionary history and future of this galaxy in order to identify the potential quenching mechanism and possible progenitors of high-redshift MQGs.

\section{{\zf} analogues}\label{sec:analogues}
In order to identify the analogues of {\zf} in our model, we consider the following two properties.
\begin{enumerate}
	\item The stellar mass of {\zf} was constrained \citep{Glazebrook2017} from the equivalent width of the observed spectral energy distribution (SED) and is considered a robust observed property. {\color{black}However, \citet{Simpson2017} recently argued that due to contaminations from a nearby dusty starburst, the stellar mass of {\zf} was overestimated by 50 per cent. Therefore, according to \citet{Simpson2017}, we set a threshold of $M_*>{\color{black}0.8}\times10^{11}\Msol$, which returns 139 massive galaxies.} Based on the Balmer absorption lines (or the lack of H$\beta$ emission line), the current SFR was reported to be less than 0.2 (or 4)$\Msol\mathrm{yr}^{-1}$. We note that in the alternate case that {\zf} is an obscured star-forming galaxy, the SFR would be less than $70 - 100 \Msol\mathrm{yr}^{-1}$, which is inferred from the non-detection threshold of the \textit{Herschel}/PACS imaging \citep{Straatman2014}. In this work, we focus on the MQG scenario and select {\zf} analogues with $\mathrm{SFR}<0.4\ \Msol\mathrm{yr}^{-1}$. This selection returns {\color{black}10 MQGs, corresponding to 7 per cent of the 139 massive galaxies.}
	
	\item\label{c2} Using spectral evolution models of stellar population \citep{1999astro.ph.12179F}, and the assumption that the stellar mass of {\zf} increases with a constant SFR for a certain period followed by a long state of quiescence, \citet{Glazebrook2017} found that {\zf} has a stellar age of $500-1050$ Myr with a formation time-scale of less than 250 Myr. This implies that {\zf} formed at $z\sim5.8^{+2.3}_{-0.8}$ with a star-forming period of $\Delta z<1.5^{+2.1}_{-0.4}$. We note that strong star formation at $z<4.5$ creates significantly excess flux in the spectrum at wavelengths shorter than $\mathrm{H}_\alpha$ compared to the observed SED of {\zf}. Therefore, we exclude galaxies with $\mathrm{SFR}>100{\Msol}\mathrm{yr}^{-1}$ at $z<4.5$ from the sample.
\end{enumerate}
Based on these specific criteria, we identify three analogues which we will henceforth refer to as ZF-1 to ZF-3. {\color{black}However, we note that the choice of the stellar mass and SFR thresholds does not have a significant impact to our conclusion.}

The {\zf} analogues in our model have stellar masses of around $10^{11}\mathrm{M}_\odot$ and negligible star formation at $z\sim3.7$ with peak SFRs $\geq500\mathrm{M}_\odot\mathrm{yr}^{-1}$ at $z\sim5-6$. We summarize their properties at $z\sim3.7$ and at the time of peak star formation in Table \ref{tab:properties}, for comparison to the {\zf} properties estimated by \citet{Glazebrook2017}. We see that, due to the limited volume of our parent \textit{N}-body simulation, the candidates are slightly less massive (by up to 30 per cent) than {\zf}. However, their halo masses (${\sim}10^{13}\Msol$) are larger than the value ($10^{12.5}\Msol$) estimated from the galaxy number density and the halo mass function at $z=4$. This implies that the three analogues have larger stellar mass to halo mass ratios by factors of $2-3$, suggesting that these MQGs are hosted in relatively large haloes.



\section{The history of {\zf}}\label{sec:history}
As discussed above, owing to the mass of {\zf} and the absence of recent star formation, there must have been a period of intense star formation at early times \citep{Glazebrook2017}. In order to identify why this star formation was subsequently quenched, we investigate the history of the three analogues by tracking their most massive progenitors along the dark matter halo merger trees.

\subsection{The role of AGN quenching in {\zf}}

\begin{figure*}
	\begin{minipage}{\textwidth}
		\centering
		\includegraphics[width=.99\textwidth]{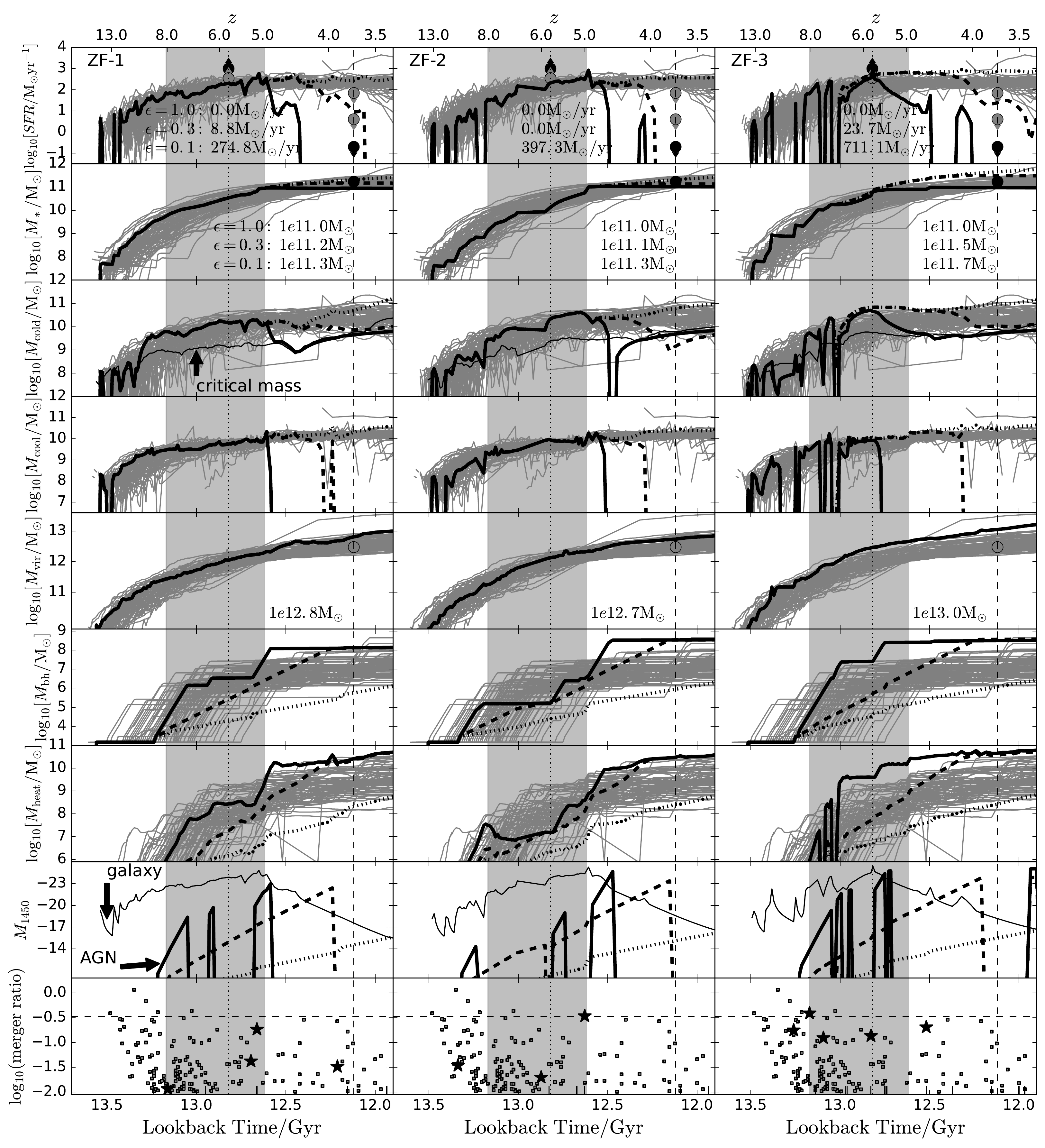}
	\end{minipage}
	\vspace*{-4mm}
	\caption{\label{fig:history} Histories of the three analogues, ZF-1, ZF-2, ZF-3, in terms of, from top to bottom, the star formation rate, stellar mass, cold gas mass, cooling mass, virial mass, central black hole mass, heating mass due to AGN feedback, intrinsic quasar UV magnitude and baryonic merger ratio. Results with different Eddington ratios, $\epsilon=1.0$, {\color{black}0.3}, 0.1, are shown with solid, dashed and dotted lines, respectively. Note that the virial masses are the same in different models and only the fiducial model with $\epsilon=1.0$ is shown in the merger ratio panels with the black star symbol. In the cold gas mass panels, thin black lines represent the critical mass in the fiducial model, above which galaxies are able to form stars. In the intrinsic quasar UV magnitude panels, thin black lines indicate the UV magnitude of the host galaxy in the fiducial model for comparison. The values of SFR, stellar mass and virial mass at $z=3.7$ are shown in each corresponding panel for the three models. The thin grey lines and small squares represent the full set of $z=3.7$ massive star-forming galaxies with $M_*>{\color{black}0.8}\times10^{11}\Msol$ and $\mathrm{SFR}>70\Msol\mathrm{yr}^{-1}$, the \textit{Herschel}/PACS non-detection threshold, in the fiducial model. The derived stellar mass and SFR limits, virial mass of {\zf} \citep{Glazebrook2017} are shown with black and grey circles. There are two SFR limits at $z\sim5.8$, representing the 68 and 95 per cent confidences and three SFR limits at $z\sim3.7$, indicating the values inferred from the Balmer absorption lines, $\mathrm{H}\beta$ emission lines and \textit{Herschel} non-detection threshold, respectively. The vertical dashed line represents the spectroscopic redshift of {\zf}, $z=3.717$. The vertical dotted line with shaded regions show the estimated time of the peak of star formation of {\zf}, $z=5.8^{+2.3}_{-0.8}$. The horizontal dashed line indicates merger ratio equal to a third. Note that merger ratios less than 1 per cent are shown with tick marks on the \textit{x}-axis for the three analogues.}
\end{figure*}

Fig. \ref{fig:history} shows the histories of the three analogues at $z\sim15-3.5$ (corresponding to a lookback time of $t\sim13.5-12.0$ Gyr). We show, from top to bottom with thick black lines and star symbols representing the fiducial model, the SFR, stellar mass, cold gas mass, cooling mass, virial mass, central black hole mass, heating mass due to AGN feedback, intrinsic quasar UV magnitude\footnote{Note that quasar UV magnitudes induced by accretion events that cannot support the quasar activity for the entire snapshot are not shown here.} and baryonic merger ratio\footnote{The baryonic merger ratio is defined as the ratio of the total mass of cold gas and stars of the merging companion to that of the target galaxy.}. In the background, with thin grey lines and squares we also plot the histories of all massive star-forming galaxies with $M_*>{\color{black}0.8}\times10^{11}\Msol$ and $\mathrm{SFR}>70\Msol\mathrm{yr}^{-1}$. The derived properties and limits of {\zf} are shown as circles for comparison. There are several points to note.
\begin{enumerate}
	\item The analogues and the massive star-forming galaxies shown in the background of Fig. \ref{fig:history} are selected at $z=3.7$. As shown in the histories of virial mass, the analogues are hosted by relatively massive haloes compared to star-forming galaxies with similar stellar masses, with masses of $M_{\rm vir}\sim 10^{13}{\Msol}$.
	
	\item Mergers trigger the most intense star formation event in the history of each of ZF-1, ZF-2 and ZF-3 at $z=5.1$, 4.9 and 5.7 with $\mathrm{SFRs}\sim600$, 800 and 1000 $\Msol\mathrm{yr}^{-1}$, and merger ratios ${\sim}0.18$, 0.34 and {\color{black}0.14}, respectively. However, unlike the star formation history modelling performed in \citet{Glazebrook2017}, where the galaxy is assumed to form stars with a constant SFR of $>990{\Msol}\mathrm{yr}^{-1}$ for only $<250$ Myr, the three analogues have longer star formation time-scales of ${\sim}500$ Myr to 1 Gyr with milder evolutions of SFR.
	
	\item After the mergers, cooling is significantly suppressed and the galaxies consume the available cold gas on short time-scales ($\sim100-300$ Myr). When there is insufficient cold gas in the galaxies ($m_\mathrm{cold}<m_\mathrm{crit}$, see equation 6 in \citealt{Mutch2016a}), star formation is quenched and the analogues become fainter in the UV band (the UV magnitude of the host galaxy is shown by the thin black line in the $M_{1450}$ panels of Fig. \ref{fig:history}).
	
	\item Mergers drive black hole growth in our model. The central massive black holes of the three galaxies therefore grow significantly after the merger event. At the time when cooling stops, their black hole masses become $1-1.5$ orders of magnitude larger than in star-forming galaxies with the same stellar mass. During the accretion phase, the central black hole radiates energy and heats the surrounding interstellar medium (ISM). In our model, AGN feedback limits gas condensation by suppressing the cooling flow \citep{croton2006many}. When the black hole becomes massive enough, AGN feedback is able to completely neutralize the cooling mass, leading to quenching of star formation in the three analogues.
	
	\item Because of the dramatic accretion following a significant merger event, the nucleus is triggered into a quasar phase \citep{Qin2017}. However, the finite accreted mass can only support central massive black hole activity for a finite period of time. In the second last row of Fig. \ref{fig:history}, we show the intrinsic quasar UV magnitude (solid thick black lines) in comparison with the magnitude of the host galaxy (solid thin black lines). We see that, before the merger, the total UV magnitude is dominated by stellar light, but that the accretion briefly dominates UV flux during the quasar phase following the merger. However later, at $z=3.7$, the nucleus has become inactive (but still provides AGN feedback through the radio mode, see more details in \citealt{Qin2017}). This explains the lack of significant UV to optical radiation from the central massive black hole in the observed SED of {\zf}, which can be well fit with a stellar population model \citep{Glazebrook2017} without AGN contributions.
\end{enumerate}

\subsubsection{Effect of the Eddington ratio}
In our fiducial model, black holes are assumed to either accrete and radiate at the Eddington rate ($\epsilon=1$, where $\epsilon$ is the Eddington ratio) or stay quiescent if there is no accretion. {\color{black}This assumption has been shown to provide a good description of black hole growth for the majority of black holes at the relevant redshifts\footnote{In our modelling we find that assuming  $\epsilon = 1$ results in an underestimate of the number of faint AGN at later times ($z < 2$) relative to observations. However, Eddington accretion with a value of $\epsilon=1$ is necessary in our model in order to produce a realistic AGN catalogue for the relevant mass range ($M_\mathrm{BH}\lesssim10^8\Msol$) and redshifts ($z > 3$) (see more in \citealt{Qin2017}).} \citep{Bonoli2009}.} In order to directly investigate the importance of the Eddington ratio in determining whether our three analogues are quenched by AGN feedback, we set the Eddington ratio to $\epsilon={\color{black}0.3}$ (dashed thick black lines in Fig. \ref{fig:history}) and $\epsilon=0.1$ (dotted thick black lines). We note that without AGN feedback, in the {\tiamat} volume the model can still reproduce the observed galaxy stellar mass function at $z>4$ but overestimates the number of more massive galaxies at $z<2$ \citep{Qin2017}. With $\epsilon=0.1$, we find that while the number of massive galaxies with $M_*>{\color{black}0.8}\times10^{11}\Msol$ increases to {\color{black}198}, the fraction of MQGs decreases significantly at high redshift (see the thin dotted line in the bottom right panel of Fig. \ref{fig:gsmf}). Moreover, amongst these galaxies, only five are quiescent galaxies. However, they cannot be considered analogues due to intense recent star formation events at $z<4.5$ that violate the selection criterion \ref{c2}. Whilst no {\zf} analogues are found in the case of sub-Eddington accretion, it is illuminating to look at the variation in the evolution of ZF-1, ZF-2 and ZF-3 with lower black hole growth efficiencies.

We see that when the central massive black hole is sub-Eddington, the black hole growth becomes slower, providing weaker feedback to galaxy formation and having a longer quasar phase with a fainter luminosity. With $\epsilon={\color{black}0.3}$, the central black holes are inactive at $z=3.7$ but the analogues form stars for a longer period down to $z<4.5$. This violates selection criterion (ii) and results in a stronger UV excess in the spectrum at wavelengths shorter than $\mathrm{H}_\alpha$ compared to {\zf}. With a lower Eddington ratio of $\epsilon=0.1$, black hole growth becomes significantly slower compared to the model with Eddington accretion, and the heating mass due to AGN becomes insignificant. We find that with $\epsilon=0.1$, the analogues are star-forming galaxies at $z=3.7$. Our model therefore predicts that the intense growth of a massive black hole ($\epsilon\sim1$) in the centre of {\zf}, likely induced by mergers, resulted in persistent AGN feedback which quenched subsequent star formation.

\begin{figure}
	\begin{minipage}{\columnwidth}
		\centering
		\includegraphics[width=1\textwidth]{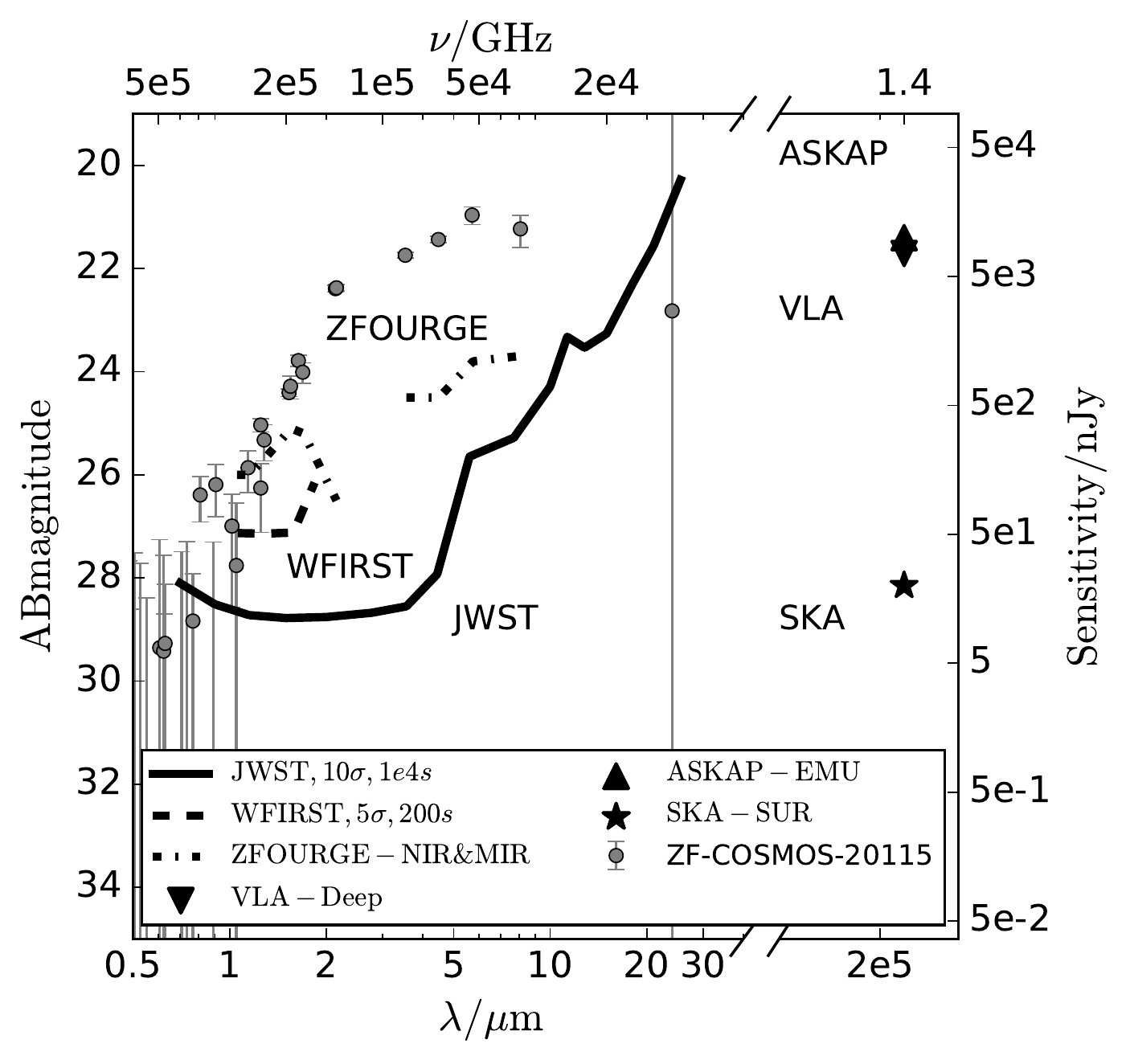}
	\end{minipage}
	\caption{\label{fig:jwst} The SED of {\zf} in the observer frame \citep{Glazebrook2017}, the sensitivities of ZFOURGE NIR and MIR instruments \citep{Dickinson2003,Straatman2016}, the expected \textit{JWST} $10\sigma$ point source detection limits \citep{Cowley2017} with ${\sim}3$h exposure time, the designed $5\sigma$ detection limits of the \textit{WFIRST} high latitude survey \citep{Spergel2013} with ${\sim}200$s exposure time and the 1.4 GHz magnitude limits of the VLA-COSMOS Deep \citep{Schinnerer2010}, ASKAP-EMU and SKA-SUR surveys \citep{2015aska.confE.173K}.}
\end{figure}

\subsubsection{Evidence for AGN feedback}
Our model suggests the existence of ${>}10^8{\Msol}$ supermassive black holes and a significant role of AGN feedback in galaxy formation at $z\sim5-6$ when the Universe was less than ${\sim}1$ Gyr old. In particular, we find that the three {\zf} analogues in our model are quenched due to persistent AGN feedback. This could be potentially examined using far-infrared (FIR) and radio observations. For instance, \citet{Gobat2017} constructed a median SED in the range of $10-10^6\mu\mathrm{m}$ using stacked images of ${\sim}1000$ MQGs at $z\sim2$. They discovered significant emissions in the FIR and radio bands of MQGs, which suggests a large content of dust (${\sim}10^8\Msol$) and gas (${\sim}10^{10}\Msol$). This is in contrast to local early-type galaxies that are usually found to be gas poor. If present, this gas must be consumed with a low efficiency in order to keep MQGs quiescent across cosmic time. This is likely due to AGN feedback and is supported by a significant excess of radio emission (${\sim}5\times10^{22}\mathrm{W/Hz}$) in the stacked SED compared to a dust only spectral model.

Conclusive evidence for this becomes observationally more challenging at $z\sim4$. We note that the near-infrared (NIR) emission of {\zf} is less than $3\mu\mathrm{Jy}$ (\textit{Spitzer}/MIPS\footnote{\url{http://www.spitzer.caltech.edu/}} 24 $\mu\mathrm{m}$) and no mid-infrared detection was reported by \textit{Herschel}/PACS\footnote{\url{http://sci.esa.int/herschel/}} (MIR, $100-160\mu\mathrm{m}$, \citealt{Straatman2015}). In addition, within a 20 arcsec box around {\zf} there is no radio source in the COSMOS VLA Deep catalogue\footnote{\url{http://irsa.ipac.caltech.edu/Missions/cosmos.html}} \citep{Schinnerer2010}, the typical rms and resolution of which are 7.5 $\mu\mathrm{Jy}$ and 1.7 arcsec. Next-generation telescopes are expected to provide enormously improved sensitivity and resolution covering a wide range of wavelengths. 

We summarize the detection capabilities of  \textit{JWST}\footnote{\url{https://jwst.nasa.gov/}}, \textit{WFIRST}\footnote{\url{https://wfirst.gsfc.nasa.gov/}}, ASKAP\footnote{\url{http://www.atnf.csiro.au/projects/askap/index.html}} and SKA\footnote{\url{http://skatelescope.org/}} in Fig. \ref{fig:jwst} compared to the SED of {\zf} in the observed frame \citep{Glazebrook2017}, the sensitivities of the NIR and MIR instruments employed by ZFOURGE \citep{Dickinson2003,Straatman2016} and the VLA-COSMOS Deep Survey. Within the context of high-redshift quiescent galaxies,  \textit{JWST} will provide broad-band imaging from 0.7 to 28.5 $\mu\mathrm{m}$ with sensitivities of ${\sim}10^{-2}\mu\mathrm{Jy}$ to $1-10\mu\mathrm{Jy}$ (3h exposure) from NIR to MIR \citep{Greene2007,2011ASPC..446..331R,Bouchet2015}. With a sufficient high-redshift MQG sample becoming available from  \textit{JWST}, radio flux can be estimated using the stacking technique with telescopes such as ASKAP \citep{Norris2011}. In addition, the high latitude survey of \textit{WFIRST} \citep{Spergel2013,Gehrels2015} will cover 6 per cent of the entire sky, a region that is 20000 times larger than ZFROUGE with AB magnitude limits of $\sim27$ in the NIR ($0.7-2\mu\mathrm{m}$). Together with the planned SKA-SUR all sky survey \citep{2015aska.confE.173K}, this will enable direct evaluation of AGN activity in MQGs.

\subsection{The possible connection between MQGs and high-redshift quasars}

\begin{figure}
	\begin{minipage}{\columnwidth}
		\centering
		\includegraphics[width=0.96\textwidth]{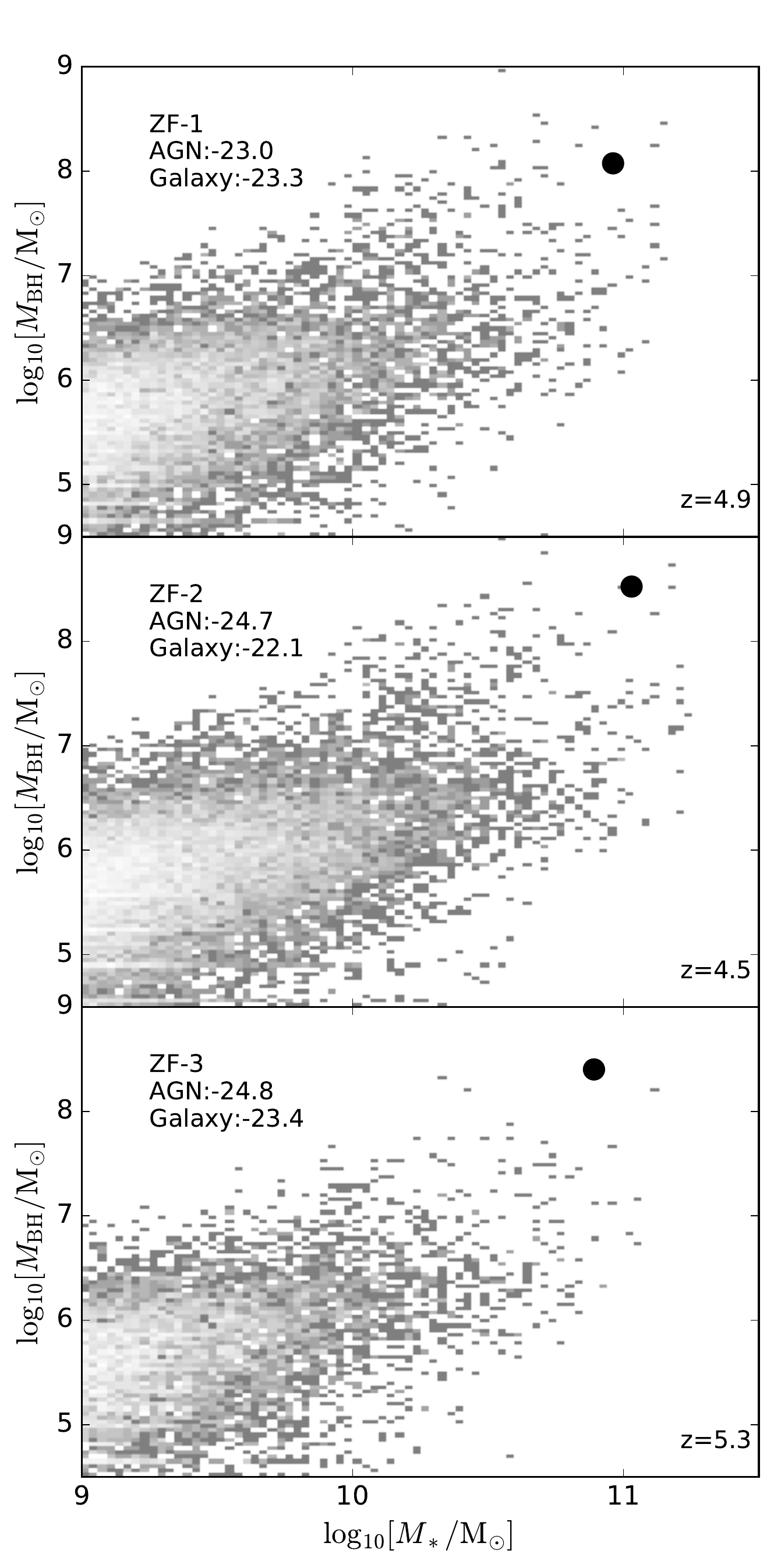}
	\end{minipage}
	\caption{\label{fig:magorrian} Correlation between black hole mass and stellar mass compared to ZF-1, ZF-2 and ZF-3 at {\color{black}$z=4.9$, 4.5} and 5.3, respectively. The three analogues at the corresponding redshift are indicated by black circles and their UV magnitudes of AGN and stellar light are shown in the top left corners.}
\end{figure}

\begin{figure}
	\begin{minipage}{\columnwidth}
		\includegraphics[width=0.97\textwidth]{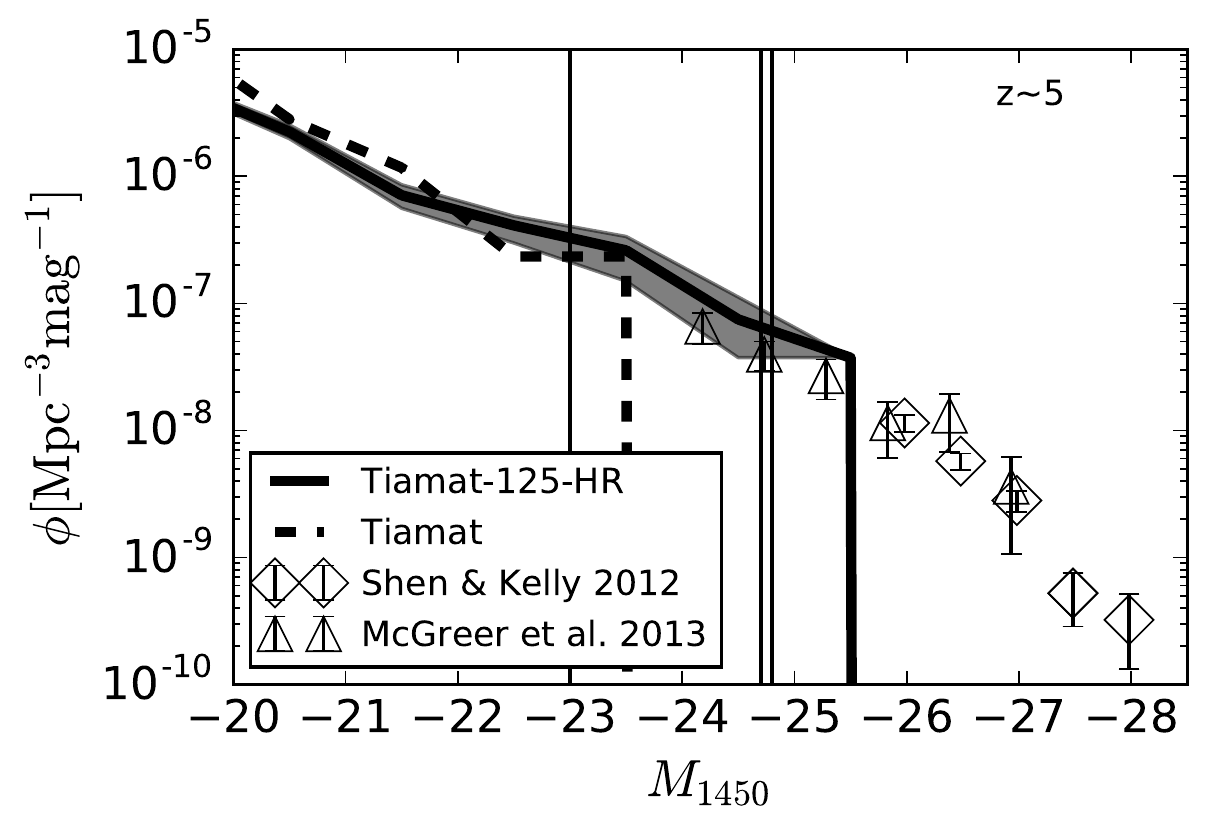}\\	\includegraphics[width=\textwidth]{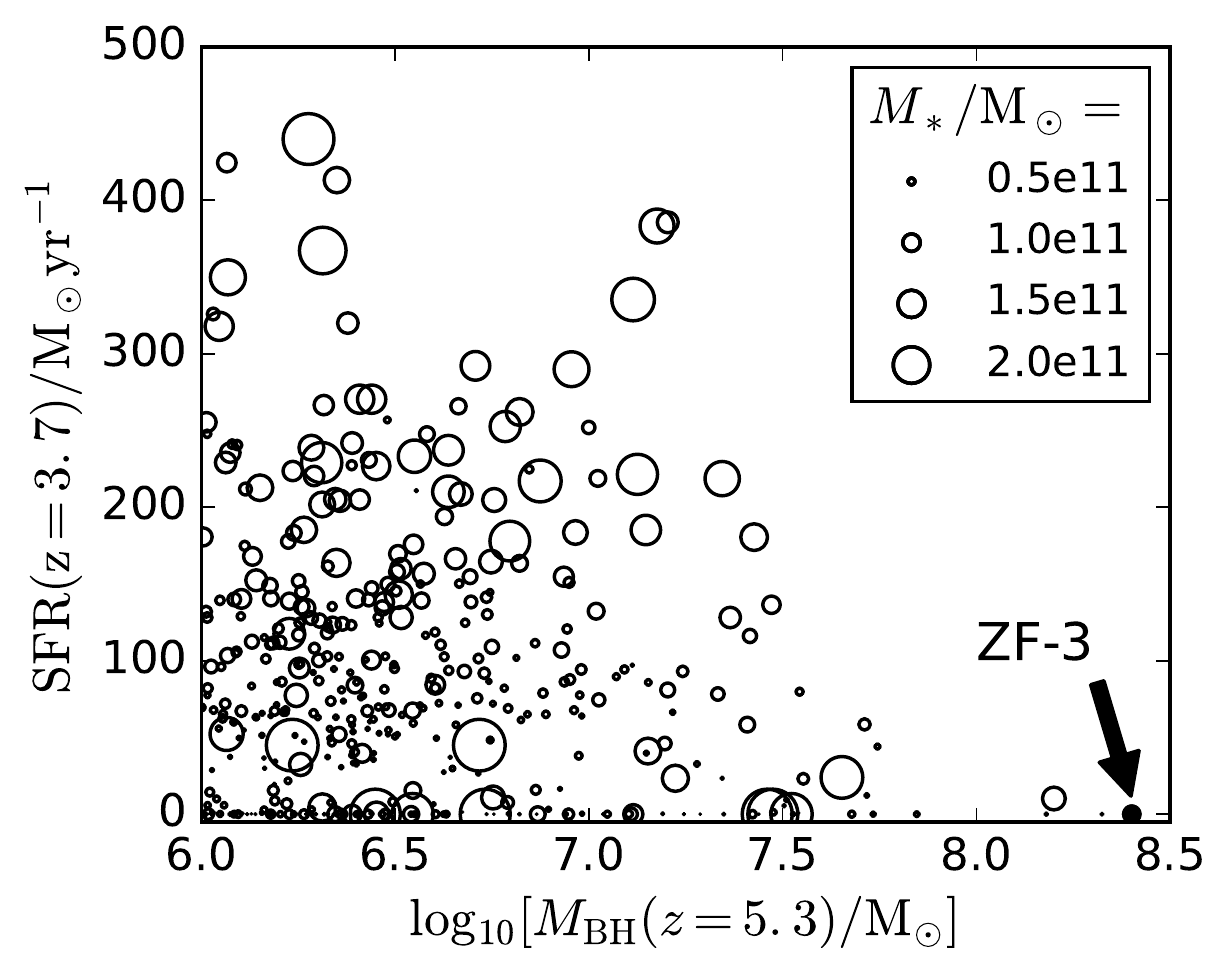}
	\end{minipage}
	\caption{\label{fig:qlf} \textit{Top panel:} quasar UV $1450\mathrm{\AA}$ luminosity function at $z\sim5$. The results using the {\tiamat} and \textit{Tiamat} halo merger trees are shown with solid and dashed lines, respectively. The shaded regions represent {\color{black}the 95 per cent confidence intervals around the mean using 100000 bootstrap re-samples} for the {\tiamat} result. The observational data is shown with different symbols: \citet{Shen2012} using SDSS DR7 data at $z\sim4.75$ ($\Diamond$) and \citet{McGreer2013} using SDSS, UKIDSS and MMT at $z{\sim}4.7-5.1$ ($\vartriangle$). From left to right, the vertical lines indicate the quasar UV magnitudes of ZF-1, ZF-2 and ZF-3, respectively. \textit{Bottom panel:} correlation between high-redshift black hole mass (z=5) and low-redshift SFR (z=3.7). The size of the empty circle represents different stellar masses and ZF-3 is indicated with the filled circle.}
\end{figure}

Fig. \ref{fig:history} shows that the UV fluxes of the three analogues for most of their histories are dominated by stellar light, and that at $z\sim3.7$ their central nuclei have become inactive. However, the central massive black holes of the three analogues have previously grown significantly (${\sim}10^8\Msol$) during the quasar phase at $z\sim5$ when accretion becomes important to or even dominates the total UV luminosity. Fig. \ref{fig:magorrian} shows the correlation between black hole mass and stellar mass of the three analogues at {\color{black}$z=4.9$, 4.5} and 5.3 when their AGN activities reached a peak. Note that the model has been shown to successfully reproduce the observed Magorrian relation at $z\lesssim0.5$ \citep{Qin2017}. We see that at $z\sim5$, the three analogues have larger black holes than other galaxies in our model, and in particular, that ZF-3 hosts the most massive black hole in the model at $z=5.3$, and has a luminosity corresponding to that of a SDSS quasar \citep{Glikman2011,Shen2012,McGreer2013}.

Indeed at the time of peak AGN activity, the quasar UV luminosities of ZF-1, ZF-2 and ZF-3 increase to {\color{black}$M_{1450}=-23.0, -24.7$ and $-24.8$, up to ${\sim}2.5$} magnitudes brighter than the host galaxies. We show the predicted quasar luminosity function at $z=5$ in the top panel of Fig. \ref{fig:qlf}. The result using the \textit{Tiamat} halo merger trees is shown for comparison and to indicate the convergence of the quasar luminosity function at $z\sim5$. The model is in agreement with observations at the bright end \citep{Shen2012,McGreer2013}. We indicate the quasar luminosities of the three analogues in the top panel of Fig. \ref{fig:qlf} and find that the progenitors of ZF-1, ZF-2 and ZF-3 are bright quasars in the range that optical surveys such as SDSS are able to detect. This suggests that {\zf} is likely the descendent of a high-redshift quasar and illustrates the possible connection between MQGs and high-redshift quasars \citep{2012MNRAS.425L..66M}. 

In order to better demonstrate this, for each massive galaxy selected at $z=5.3$, we present the correlation between black hole mass and subsequent SFR at $z=3.7$ in the bottom panel of Fig. \ref{fig:qlf}. ZF-3 is indicated with a filled circle. We see that galaxies with less massive black holes ($M_\mathrm{BH}<10^{7.5}\Msol$) can have a range of subsequent star formation levels. This is due to self-regulated stellar feedback being the dominant mechanism in these galaxies. However, due to persistent AGN feedback, galaxies with more massive black holes do not have intense subsequent star formation. This indicates that a high-redshift bright quasar will likely become a MQG at later times, when its accretion disc has been completely consumed.

\subsection{The reionization history}

\begin{figure}
	\begin{minipage}{\columnwidth}
		\centering
		\includegraphics[width=\textwidth]{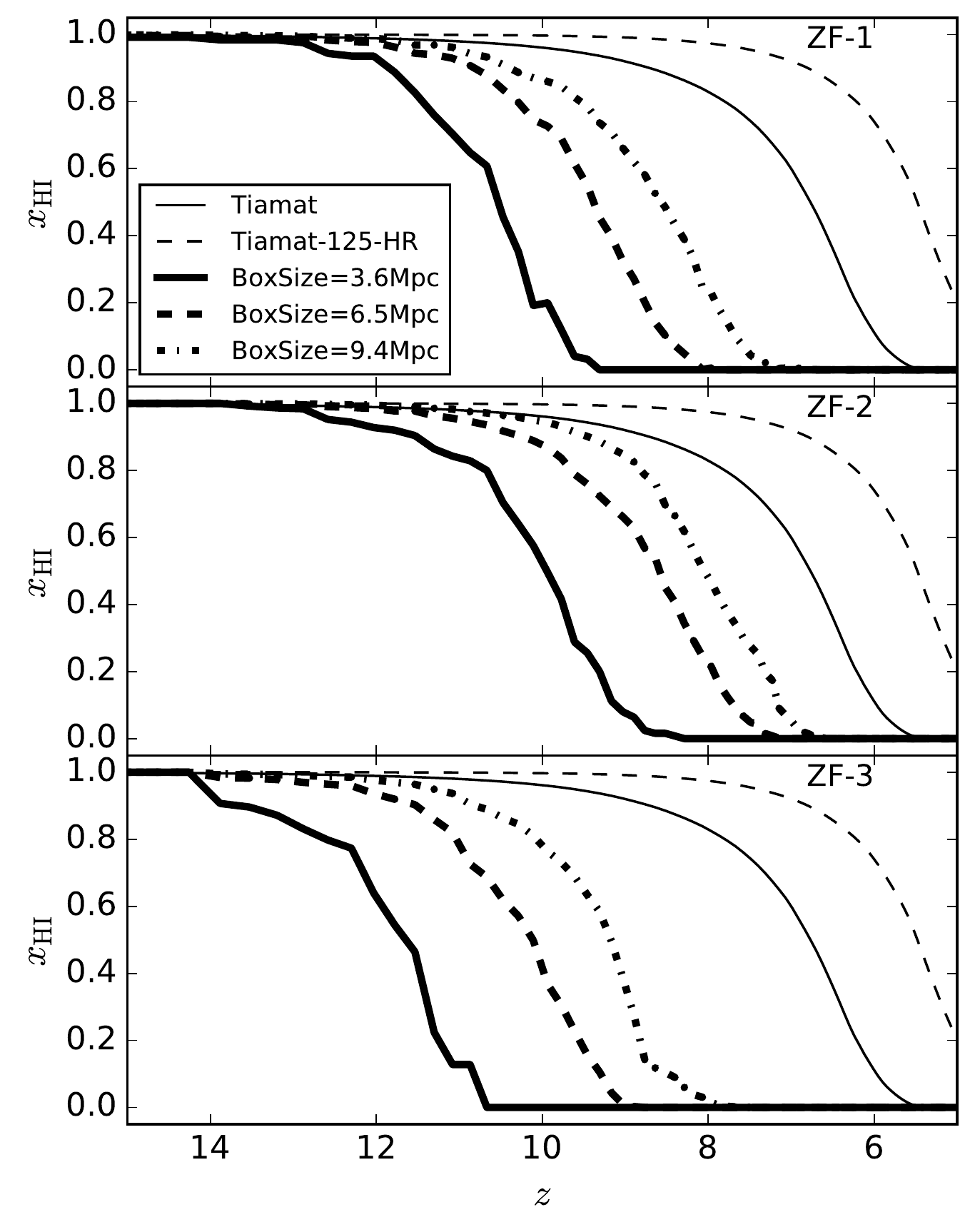}
	\end{minipage}
	\caption{\label{fig:reionization} The evolution of the average neutral hydrogen fraction within {\color{black}boxes of lengths around 3.6, 6.5 and 9.4 Mpc} around the three {\zf} analogues. The result within the entire {\tiamat} and \textit{Tiamat} volumes are shown for comparison.}
\end{figure}

The three {\zf} analogues are hosted by massive haloes, which provide deep gravitational potentials that efficiently accrete baryons. The high baryon accretion efficiency induces intense star formation at high redshift, providing a significant number of ionizing photons that reionize surrounding hydrogen at a very early time. In this section, we discuss the reionization history of {\zf}.

\begin{figure*}
	\begin{minipage}{\textwidth}
		\centering
		\includegraphics[width=1.02\textwidth]{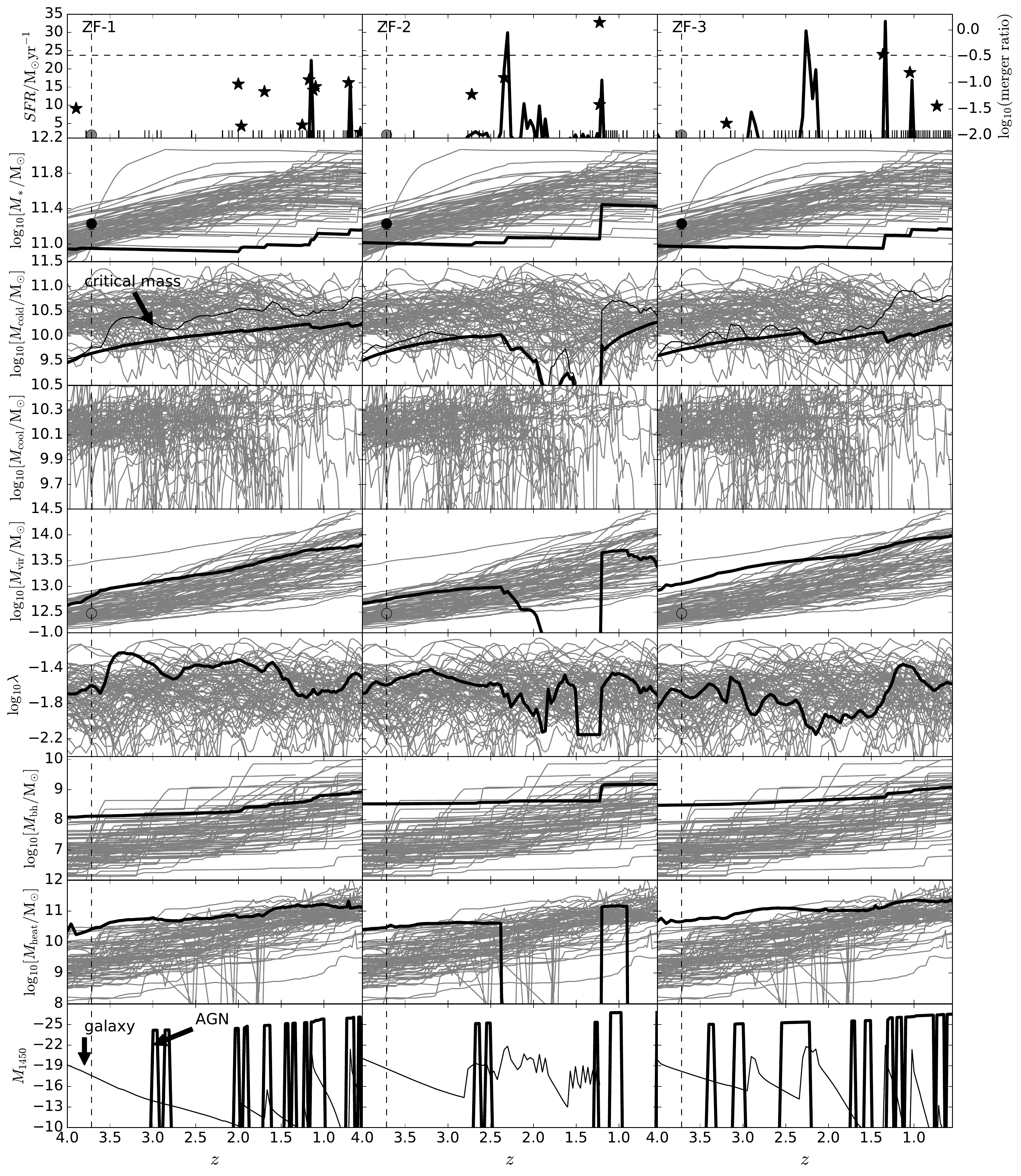}
	\end{minipage}
	\caption{\label{fig:future} Futures of the three analogues, ZF-1, ZF-2, ZF-3, in terms of, from top to bottom, the star formation rate (and the baryonic merger ratio on the right \textit{y}-axis), stellar mass, cold gas mass, cooling mass, virial mass, halo spin parameter, central black hole mass, heating mass due to AGN feedback and intrinsic quasar UV magnitude. In the cold gas mass panels, thin black lines represent the critical mass, above which galaxies are able to form stars. In the intrinsic quasar UV magnitude panels, thin black lines indicate the UV magnitude of the host galaxy for comparison. Thin grey lines represent the full sample of massive star-forming galaxies with $M_*>{\color{black}0.8}\times10^{11}\Msol$ and $\mathrm{SFR}>70\Msol\mathrm{yr}^{-1}$, the \textit{Herschel}/PACS non-detection threshold. The derived stellar mass, virial mass and SFR limits of {\zf} \citep{Glazebrook2017} are shown with black and grey circles. The vertical dashed line represents the spectroscopic redshift of {\zf}, $z=3.717$. The horizontal dashed line indicates merger ratio equal to a third. Note that merger ratios less than 1 per cent are shown with tick marks on the \textit{x}-axis for the three analogues.}
\end{figure*}

In Fig. \ref{fig:reionization}, we show the evolution of the average neutral hydrogen fraction within boxes with different sizes around the three analogues. The neutral hydrogen fraction is calculated using {\tocf} \citep{Mesinger2011} with the galaxy catalogue provided by {\meraxes} (see more details in \citealt{Mutch2016a}). We see that the analogues start ionizing the intergalactic medium (IGM) at $z\gtrsim14$ with inner regions becoming ionized first. For example, ZF-1, ZF-2 and ZF-3 (and their nearby galaxies) reionize {\color{black}3.6Mpc regions at $z\sim 9.5$, 9 and 10.5, respectively, and 9.4Mpc regions by $z\sim7$, 6.5 and 8}. We also present the average neutral hydrogen fraction of the entire {\tiamat} box in Fig. \ref{fig:reionization}.

Due to the simulation not resolving dwarf galaxies, reionization only finishes at $z<5$ in the {\tiamat} volume. We therefore also show the result calculated using the higher resolution \textit{Tiamat} halo merger trees for comparison. This higher resolution model has been shown \citep{Qin2017} to match the observed ionizing emissivity at $z\sim5-2$ \citep{Becker2013} and the Thomson scattering optical depth \citep{PlanckCollaboration2016}. Comparing the average evolution in ionized fraction between the two simulations illustrates that reionization predicted are delayed by $\Delta z\sim1$ in the low-resolution simulation. The model suggests that {\zf} is located in a region of IGM that was ionized early in the reionization era.

\section{the subsequent evolution of {\zf}}\label{sec:future}

In this section, we use the {\meraxes} semi-analytic model to explore possible scenarios for the subsequent evolution of {\zf} at $z<3.7$. Fig. \ref{fig:future} shows the future of the three analogues from $z\sim4$ to $z\sim{\color{black}0.56}$. We see that AGN feedback keeps gas in the analogues from cooling, and that without a major merger it is unlikely that a galaxy like {\zf} will have a further strong star formation event at $z<3.7$. However, depending on the halo properties or merger ratios, starbursts can in some cases be triggered. Because {\zf} is likely to be cold gas poor due to AGN heating, subsequent starburst events would only persist for a short time period. We also see the following trends.

\begin{enumerate}
	\item When there is a merger, the three analogues increase their stellar masses by absorbing baryons from their merging companions. The UV flux changes due to different star-forming histories of the merging companion (most likely the companion is a star-forming galaxy and therefore the UV flux increases after the merger), as well as the merger-triggered starburst. However, depending on the merger ratio, only a small fraction of mergers can trigger additional large starbursts. For example, ZF-1 stays quiescent until $z\sim1.1$ when several galaxies merge with it, inducing a minor star-forming event with a $\mathrm{SFR}\sim25{\Msol}\mathrm{yr}^{-1}$.
	
	\item With the central supermassive black hole heating the ISM, continuous star formation is unlikely to occur. However, there are three short ($\Delta z<0.1$) star formation events in ZF-3 at $z\sim2-3$, when the cold gas mass, which gradually increases from stellar mass recycling \citep{Mutch2016a}, exceeds the critical value. We note that the star formation law in our model depends on a critical mass, above which galaxies are able to form stars. This critical mass is inferred from the host halo properties and increases with virial mass and halo angular momentum (based on \citealt{kennicutt1998global} and \citealt{kauffmann1996disc}). While the halo mass of ZF-3 grows smoothly, the spin parameter, which is a measure of specific angular momentum, decreases by a factor of 2 during the star-forming events. Assuming full conservation of specific angular momentum for newly cooling gas \citep{Mutch2016a}, the cold gas disc contracts, resulting in higher gas densities. Under circumstances where the cold gas density becomes larger than the critical value, star formation can occur.
	
	\item ZF-2 merges\footnote{Note that using a central weighting algorithm, the {\tiamat} halo merger trees were constructed by tracking halo cores instead of the majority of halo particles. Since the core is where a galaxy forms and evolves, this strategy improves the semi-analytic galaxy formation modelling. In this work, {\sc subfind} \citep{springel2008aquarius}, a standard halo finder, is adopted, and in some circumstances {\sc subfind} redistributes the dark matter particles of a central halo to its satellites and leads to a central halo not being the most massive halo (central-satellite switching, see more details in Poole et al. in preparation). This results in unrealistic changes in halo properties. For clarity, we do not show the properties when this occurs and we note that this does not have any impact on our conclusions or predictions of the future of {\zf}.} with a larger galaxy at $z\sim2.5-1.2$. The host halo of ZF-2 initially becomes smaller due to tidal stripping from the merging companion, and eventually merges into the more massive halo with a baryonic merger ratio of ${\sim}1.4$. When the halo is being stripped, the host galaxy loses its gas component which becomes unbound. Consequently, without gas falling into the centre, the central massive black hole stops accreting and there is no AGN feedback or heating in ZF-2 during the stripping period. On the other hand, the cold gas disc is considered to be gravitationally bound in our model and remains intact during galaxy stripping. When the halo mass becomes smaller, the disc size and hence the critical mass required for star formation decreases. This results in star formation activity which quickly consumes the available gas.
\end{enumerate}

\section{conclusions}\label{sec:conclusions}
{\zf} is a MQG at $z\sim3.7$. We model MQGs using the {\meraxes} semi-analytic model and identify three {\zf} analogues in the fiducial model presented in \citet{Qin2017}. These analogues have properties that are in agreement with the observed constraints on {\zf} inferred from the recent spectroscopic follow-up of \citet{Glazebrook2017}. 

We find that the three analogues are hosted by more massive haloes ($10^{13}\Msol$) compared to other star-forming galaxies with similar stellar masses. Following the most massive progenitor, we track the history of the three analogues and identify significant merger events at $z\sim5$. We find that when the mergers drive intense Eddington-limited growth of the central massive black hole, the cooling flow is significantly suppressed by the resulting AGN feedback. We further investigate scenarios when black holes accrete at sub-Eddington rates and find insufficient heating energy from AGN feedback to suppress star formation. In particular, with black holes accreting at {\color{black}a third} of the Eddington rate, the {\zf} analogues still have high SFRs at $z\sim4.5$, while, with the Eddington ratio of 0.1, they are star-forming galaxies at $z\sim3.7$. Our model therefore suggests that there was a period when the central massive black hole grew rapidly in {\zf}, probably triggered by mergers, and that the persistent feedback from AGN quenched the subsequent star formation. 

In addition, we find the three analogues host the most massive black holes in our simulation, and that they were luminous quasars ($M_{1450}\lesssim-23$) at $z\sim5$. Moreover, all galaxies with massive black holes at $z\sim5$ have quenched star formation by $z\sim3.7$. This suggests that {\zf} is the descendent of a high-redshift quasars similar to those detected in SDSS. By investigating the ionizing regions around the three analogues, we find that {\zf} formed in a region that was reionized early. We also follow the future of our {\zf} analogues down to $z={\color{black}0.56}$ and find that {\color{black} further strong or continuous star formation events are unlikely to occur in {\zf} at $z<3.7$}.

\section*{Acknowledgements}
This research was supported by the Victorian Life Sciences Computation Initiative (VLSCI), grant ref. UOM0005, on its Peak Computing Facility hosted at the University of Melbourne, an initiative of the Victorian Government, Australia. Part of this work was performed on the gSTAR national facility at Swinburne University of Technology. gSTAR is funded by Swinburne and the Australian Governments Education Investment Fund. This work was supported by the Flagship Allocation Scheme of the NCI National Facility at the ANU, generous allocations of time through the iVEC Partner Share and Australian Supercomputer Time Allocation Committee. AM acknowledges support from the European Research Council (ERC) under the European Union's Horizon 2020 research and innovation program (Grant No. 638809 -- AIDA).
\bibliographystyle{\dir mn2e}
\bibliography{reference}

\appendix
\label{lastpage}
\end{document}